\documentclass[prl,twocolumn,aps,showpacs,amsmath,amssymb]{revtex4}
\usepackage{graphicx}
\usepackage{dcolumn}
\usepackage{bm}

\begin{document}

\title{Stability of half quantum vortex in rotating superfluid
$^3$He-A between parallel plates}

\author{T. Kawakami}
\affiliation{Department of Physics, Okayama University,
Okayama 700-8530, Japan}
\author{Y. Tsutsumi}
\affiliation{Department of Physics, Okayama University,
Okayama 700-8530, Japan}
\author{K. Machida}
\affiliation{Department of Physics, Okayama University,
Okayama 700-8530, Japan}
\date{\today}

\begin{abstract}
We have found the precise stability region of the half quantum vortex (HQV)
for superfluid $^3$He A phase confined in parallel plates with a narrow gap
under rotation. Standard Ginzburg-Landau free energy, which is well established,
is solved to locate the stability region spanned by
temperature $T$ and rotation speed ($\Omega$).
This $\Omega$-$T$ stability region is wide enough to check it experimentally
in available experimental setup.
The detailed order parameter structure of HQV characterized by A$_1$ core is given to
facilitate the physical reasons of its stability over other vortices or textures.
\end{abstract}


\maketitle

Half quantum vortex (HQV) and associated Mojorana zero energy mode
have been widely discussed in various
research fields in condensed matter physics, ranging from
superconductors, superfluids, graphene and fractional Quantum Hall
systems\cite{general}.
In particular, theoretical
and experimental investigations are devoted
to finding HQV in superconductors and neutral Fermion superfluids
in cold atoms.
Recently strong interest on HQV is partly motivated
by the fact that the bound state created in the core of
HQV is characterized by the Majorana state with the
zero energy exactly at the Fermi level.
The Majorana particle\cite{majorana} is thought to be a candidate
for quantum computation because it obeys non-Abelian statistics\cite{ivanov}
and its existence is protected topologically to avoid decoherence.
These situations are ideal for quantum computation\cite{dassarma}
if it really exists.

So far there has been no firm experimental evidence for HQV in
any superconductors. It is necessary for HQV to exist that superconductivity
is described by a chiral $p$-wave pairing where the {$\vec d$} vector
is able to be free to rotate. It has often been argued that Sr$_2$RuO$_4$
may be a prime candidate\cite{maki,sarma,chung}, but strong doubt has been cast on this possibility of Sr$_2$RuO$_4$ of 
its triplet pairing\cite{machida2,lebed,mazin}.
Note that the first discovered triplet superconductor UPt$_3$ is an
$f$-wave pairing, not chiral $p$-wave\cite{machida1}.

Superfluid $^3$He-A phase is characterized by a
chiral $p$-wave pairing.
There is no doubt on this identification\cite{leggett}. In fact, Volovik and Mineev\cite{mineev}
are the first to point out the possibility to
the realization of HQV in 1976. Since then, there have been
several general arguments on the stability of a HQV in connection with
$^3$He-A phase\cite{salomaa,salomaaRMP,volovikbook}. However, there are no
serious calculations which consider realistic situation in
superfluid $^3$He-A phase on how to stabilize it and on what boundary conditions are needed for it.

Recently, Yamashita, et al\cite{yamashita} have performed an experiment intended
to observe HQV in superfluid $^3$He-A in parallel plate geometry.
The superfluid is confined in a cylindrical region with the radius $R=1.5$mm
and the height 12.5$\mu$m sandwiched by parallel plates.
A magnetic field $H=26.7$mT($\parallel$$z$) is applied
perpendicular to the parallel plates
under pressure $P$=3.05MPa.
Since the gap 12.5$\mu$m
between plates is narrow compared to the dipole coherence length
$\xi_d\sim 10\mu$m, the $\vec l$ vector is always perpendicular to the plates.
Also the $\vec d$ vector is confined within the plane because
the dipole magnetic field $H_d\sim 2.0$mT.
They investigate to seek out various parameter spaces, such as
temperature $T$, or the rotation speed $\Omega$ up to $\Omega=6.28$rad/s
by using the rotating cryostat in ISSP, Univ. Tokyo, capable for the maximum rotation speed $\sim$12rad/s,
but there is no evidence for HQV\cite{yamashita}.
Here we are going to give an answer why it is so
and to examine the stability region of a HQV
which competes with the ordinary singular vortex with the integer
winding number and propose a concrete experimental setup which is feasible
to perform in the light of the present experimental situation\cite{yamashita}.

We start by examining the possible order parameter (OP)
forms allowed under the above
experimental conditions. The OP of the superfluid $^3$He is given by
${\hat \Delta}=i\Sigma_{\mu,i}(A_{\mu,i}\sigma_{\mu}{\hat p}_i)\sigma_y$
in general where $A_{\mu,i}$ is a 3$\times$3 matrix $(\mu,i=x,y,z)$.
Among the two known bulk phases as ABM (A) and BW (B) phases,
we focus on the A phase in this paper, which is written as
$A_{\mu,i}=\Delta_0d_{\mu}(\vec {n}+i\vec {m})$ conveniently
expressed in terms of the $\vec d$ vector
and the $\vec l$ vector ($\vec l=\vec n\times \vec m$, $\vec n$ and $\vec m$
are unit vectors forming a triad). The former (latter) vector characterizes
the spin (orbital) part of the OP. Since the $\vec l$ vector is locked
perpendicular to the plates along the $z$ axis and the applied field ($H\parallel z$)
perpendicular to the plates confines the $\vec d$ vector  within the
plane $(x,y)$ because it is strong enough compared with the dipole field $H_d$
as mentioned above.

The HQV form originally proposed by Salomaa and
Volovik\cite{salomaa}, which was followed by others\cite{maki,chung,ivanov},
can be expressed as

\begin{eqnarray}
A_{\mu,i}=\Delta_0 e^{i\theta/2}d_\mu(\vec n+i\vec m)_i
\end{eqnarray}

\noindent
where $\vec {d}=\hat {x}\cos{\theta\over 2}+\hat {y}\sin{\theta\over 2}$
($\theta$ is the angle from the $\hat x$ axis). Since the $\vec d$ vector
is assumed to be real here, we call it R-HQV.
When winding around the vortex core, the OP exhibits simultaneous change
of sign of the $\vec d$ vector and shift of the phase $\theta$. Namely,
$(\theta, \vec d)\Longrightarrow (\theta+\pi, -\vec d)$.
The $\pi$ phase windings of the orbital and the spin parts add up,
resulting in $2\pi$ phase winding in total.
Alternatively, the wave function of this R-HQV form is cast in a form

\begin{eqnarray}
\psi=\Delta_0(r)(e^{i\theta}|\uparrow\uparrow\rangle+|\downarrow\downarrow\rangle)
(p_x+ip_y).
\end{eqnarray}

\noindent
It is clear that in the R-HQV the $\uparrow\uparrow$ pairs phase-wind by 2$\pi$
while the $\downarrow\downarrow$ pairs do not.
It will turn out shortly that this somewhat restrictive R-HQV form
is not a full solution of our Ginzburg-Landau (GL) free energy functional
under rotation.
Thus we have to seek more general HQV solution to be competitive with
the vortex free state stable at rest (A phase texture; AT)
and the singular vortex with integer winding (SV).
We generalize the OP to find the stable HQV by noticing
that the orbital part is doubly degenerate $p\pm ip_y=p_{\pm}$
in addition to doubly degenerate spin space. The most general
wave function is spanned by four basis functions, namely,

\begin{eqnarray}
\psi=\bigl(A_{++}(r,\theta)|\uparrow\uparrow\rangle
+A_{-+}(r,\theta)|\downarrow\downarrow\rangle\bigr)p_+\nonumber\\
+\bigl(A_{+-}(r,\theta)|\uparrow\uparrow\rangle
+A_{--}(r,\theta)|\downarrow\downarrow\rangle\bigr)p_-
\end{eqnarray}

\noindent
where each component
$A_{\pm\pm}(r,\theta)=A_{\pm\pm}(r)e^{i\theta w_{\pm \pm}}$
can have its own winding number $w_{\pm\pm}$ in the polar coordinates.
Under axis-symmetry $w_{\pm +}=w_{\pm -}-2n$ must be satisfied
with $n>0$ being integer\cite{isoshima}.
The winding number combination $(w_{++},w_{-+},w_{+-},w_{--})=(1,0,3,2)$
is straightforwardly generalized from the above R-HQV form (2),  which
we are trying to stabilize. The (0,0,2,2) phase is the A phase texture (AT)
and (1,1,3,3) is the ordinary singular vortex (SV).
The AT is always stable at rest, and HQV and SV compete each other
under rotation. Other several phases with different
winding number combinations, such
as (0,-1,2,1), (-1,-2,1,0) or (-1,-1,1,1) are all irrelevant; namely they are never stabilized. Note that the (0,-1,2,1) phase is stable next to the lowest AT (0,0,2,2)
at rest. Under counter clock-wise rotation the chiral $p_+=p_x+ip_y$ is favored
over $p_-=p_x-ip_y$. Thus the $p_+(p_-)$ component constitutes the
major (minor) one.

The Ginzburg-Landau (GL) free energy functional is well established\cite{leggett,salomaaRMP,sauls,greywall,kita} and given by a standard form


\begin{eqnarray}
f_{total}=f_{grad}+f_{bulk}+f_{dipole}
\end{eqnarray}

\begin{eqnarray}
f_{grad}&=& K [(\partial_i^* A_{\mu j}^*)(\partial_i A_{\mu j}) + (\partial_i^* A_{\mu j}^*)(\partial_j A_{\mu i})\nonumber\\
&+&(\partial_i^* A_{\mu i}^*)(\partial_j A_{\mu j} )]
\end{eqnarray}

\begin{eqnarray}
f_{bulk}&=&-\alpha_{\mu} A_{\mu i}^* A_{\mu i} +\beta_1 A_{\mu i}^* A_{\mu i}^* A_{\nu j} A_{\nu j}\nonumber\\
&+&\beta_2 A_{\mu i}^* A_{\nu j}^* A_{\mu i} A_{\nu j}
+\beta_3 A_{\mu i}^* A_{\nu i}^* A_{\mu j} A_{\nu j}  \nonumber\\
&+&\beta_4 A_{\mu i}^* A_{\nu j}^* A_{\mu j} A_{\nu i}
+\beta_5 A_{\mu i}^* A_{\mu j}^* A_{\nu i} A_{\nu j}
\end{eqnarray}

\begin{eqnarray}
f_{dipole}&=& g_d ( A_{\mu \mu}^* A_{\nu \nu} +A_{\mu \nu}^* A_{\nu \mu}
-{2\over 3}A_{\mu \nu}^* A_{\mu \nu})
\end{eqnarray}

\noindent
where
$\partial_i= \nabla_i -i\frac{2m_3}{\hbar }({\vec \Omega \times \vec r})_i$ $(\vec\Omega//z)$,
$\alpha_{\mu} = \alpha_0\left( 1 - T/T_c +\mu \Delta T/T_c\right)$ ($\mu=\pm$, $\alpha_0=\frac{N(0)}{3}$),
$K={7 \zeta (3) N(0) ( \hbar v_F )^2}/{240 (\pi k_B T_c)^2}$. 
$g_d$ is  the coupling constant of the dipole interaction, which is $g_d\ll\alpha_0$\cite{leggett}.
As mentioned above, we assume a two-dimensional system for the OP
spatial variation; $\mu,i= x, y$ or $\pm$.
The magnetic field acts not only to pin the $\vec d$ vector within the plane,
but also to shift the transition temperature $T_{c}$ by $\Delta t=\Delta T/T_c
=(T_{c\downarrow}-T_{c\uparrow})/2T_c$.
The fourth order GL coefficients are given by
$\beta_1 =-(1+0.1\delta)\beta_0$,
$\beta_2 =(2+0.2\delta)\beta_0$,
$\beta_3 =(2-0.05\delta)\beta_0$,
$\beta_4 =-(2-0.055\delta)\beta_0$, and
$\beta_5 =-(2+0.7\delta)\beta_0$
where
$\beta_0 ={7\zeta (3)N(0)}/{120 \left( \pi k_B T_c \right)^2}$\cite{salomaaRMP}.
The strong coupling correction $\delta > 0$ due to spin fluctuations
serves stabilizing the A phase over the B phase in the $(P,T)$
phase diagram\cite{anderson}. In the following we use the GL parameters\cite{GLparameters} tabulated\cite{sauls,greywall,kita} appropriate for the
experiment at $P$=3.05MPa.

We find the free energy minima under the rigid boundary condition
$A_{\mu i}=0$ for $r\geq R$ ($R$ is the radius of the system).
A fundamental difficulty associated with the numerical computations lies in
the fact that the coherent length $\xi=10$nm is extremely small compared with the system size $R$ where we have to take care of these two
length scales in the equal footing in order to accurately evaluate the
relative stability among three textures; AT, HQV and SV. This is a
reason why this kind of serious energy comparison has not been done before.
We carefully calibrate the accuracy of our numerical computation to
allow the detailed comparison.

\begin{figure}
\includegraphics[width=6cm]{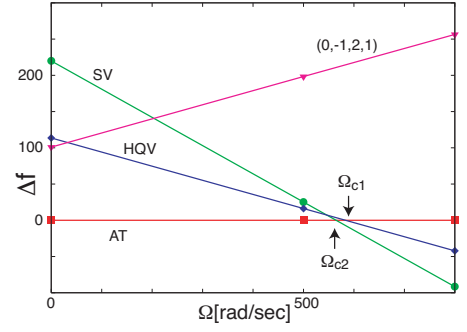}
\caption{(Color online)
Free energy comparison for AT, HQV, SV and (0,-1,2,1) states
as a function of $\Omega$ for $R=10\mu$m and $t=T/T_c=0.95$.
}
\end{figure}

We first consider the weak field case where the transition temperature
splitting $\Delta t\simeq 0$.
As shown in Fig. 1 we compare three phases with additional other phase
mentioned above. It is seen that at rest and lower rotation region
AT is stable and eventually upon increasing $\Omega$,
SV takes over at $\Omega_{c2}$. Although the HQV is stabler than SV at rest
situated almost at the half way between AT and SV
because the phase winding occurs only for the $\uparrow\uparrow$ pairs.
Under rotation the energy gain due to the angular momentum is less than
that in SV because of the above reason. Thus HQV is never
stabilized under weak field region. We also plot the (0,-1,2,1) state which is
second lowest at rest and becomes irrelevant under rotation.

\begin{figure}
\includegraphics[width=7cm]{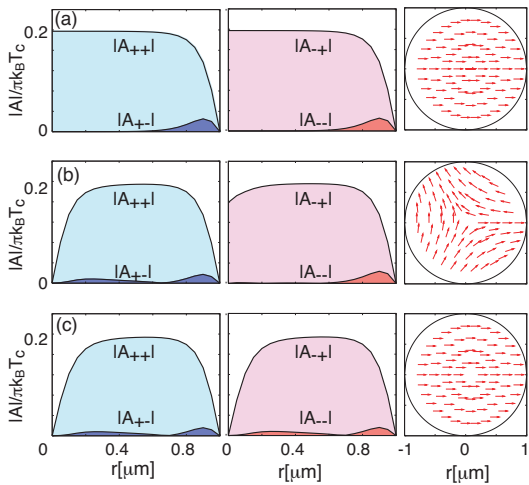}
\caption{(Color online)
Order parameter and $\vec d$ vector profiles for AT (a),
HQV (b) and SV (c) for $R=1.0\mu$m and $t=0.97$.
Left and center columns show the cross-section of OP along
the radial direction $r$. Right column shows $\vec d$ vector
patterns. In (b) it winds by $\pi$ around  the center $r=0$.
}
\end{figure}

Note that the previous form (2) of R-HQV does not improve this situation,
rather becomes worse in its stability. The R-HQV (2) indicates that
the vortex core singularity occurs for both $\uparrow\uparrow$
and $\downarrow\downarrow$ pairs even though the latter does not
have phase-winding, leading to the additional loss of the
condensation energy. The strong coupling effect acts to destabilize
both R-HQV and SV relative to AT, thus R-HQV is never stabilized (see below).

In Fig. 2 we illustrate the results of the OP profiles for
AT (a), HQV (b) and SV (c) and their $\vec d$ vector textures.
The OP's in AT are uniform in the central region around $r=0$,
decreasing towards zero at the boundary $r=R$ whose characteristic
length is $\xi$. Thus AT is basically the A phase in the bulk.
Near the boundary the induced components $A_{+-}$ and $A_{--}$ appear peripherally.

In HQV (Fig.2(b)) one of the two majority components $A_{++}$ with $w_{++}=1$
exhibits a phase singularity at $r=0$, the other component
$A_{-+}$ with $w_{-+}=0$ being depressed slightly there.
$A_{+-}$ with $w_{+-}=3$ and $A_{--}$ with $w_{--}=2$
are also induced at $r=0$ and  $r=R$. Therefore this HQV profile
shows that only the $\downarrow\downarrow$ pairs appears at around $r=0$,
implying the $A_1$ core state. This tends to stabilize this HQV further compared with R-HQV given by Eq.(2)
because (A) the condensation energy loss is less,
(B) the fourth order GL energy concerning the interaction term between
$\uparrow\uparrow$ and $\downarrow\downarrow$ pairs can be expressed as
$-4\delta\beta_0|d_+|^2|d_-|^2$.
This particular term due to the strong coupling acts to earn the extra gain
for this HQV. However, AT is simultaneously stabilized by this term, thus
HQV never wins in weak fields. Note in passing that the $\uparrow\uparrow$ and
$\downarrow\downarrow$ pairs are
completely independent when $\delta=0$
because the weak coupling GL form is derived under
the assumption that the spin space is rotationally invariant.
It is seen from Fig.2 that the $\vec d$ vector rotates
by $\pi$ when going around the origin in HQV (b) while
in the others (a) and (c) it is uniform.

Finally SV in Fig.2(c) exhibits the phase singularities for both major components
$A_{++}$ with $w_{++}=1$ and $A_{-+}$ with $w_{-+}=1$ and the induced components
$A_{+-}$ with $w_{+-}=3$ and $A_{--}$ with $w_{--}=3$  appear
at the places where the OP spatially varies.
Thus this SV is quite advantageous under rotation because
they can absorb efficiently the rotational kinetic energy.

\begin{figure}
\includegraphics[width=7cm]{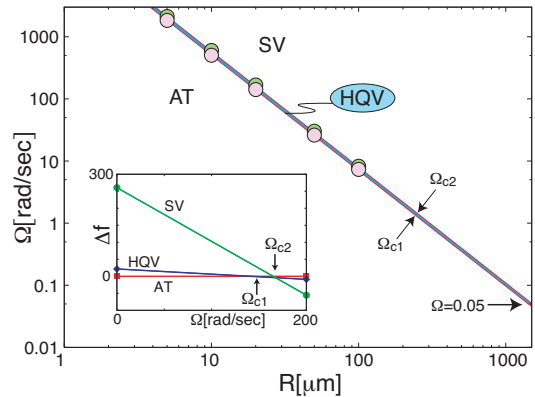}
\caption{(Color online)
Stability region of HQV sandwiched between $\Omega_{c1}$
and $\Omega_{c2}$ as a function of $R$ ($t=0.97$ and $\Delta t=0.05$).
$\Omega_{c1}$=0.05 rad/sec is the extrapolated value for $R$=1.5mm.
Inset shows the free energy comparison for $R=20\mu$m,
displaying the successive transitions from AT to HQV at $\Omega_{c1}$
and from HQV to SV at $\Omega_{c2}$.
}
\end{figure}

Having found that HQV is not stable in weak field region ($H \sim H_d=$10mT)
both at rest and under rotation, we resort to higher field region;
an order of a few kG where $\Delta t\neq 0$ or $T_{c\uparrow}\neq T_{c\downarrow}$.
This extension indeed stabilizes the HQV as shown in the inset
of Fig. 3 where we compare the three states as a function of $\Omega$.
It is seen that as increasing $\Omega$, AT changes into HQV at $\Omega_{c1}$
and then HQV to SV at $\Omega_{c2}$. The relative stability region
$\Omega_{c1}/\Omega_{c2}\sim0.8$ which is wide enough to check experimentally.

The reason for the HQV stabilization is physically explained as follows.
(A) By introducing $\Delta t$ which increases (decreases) the OP amplitude
of $\downarrow\downarrow$ pair ($\uparrow\uparrow$ pair),
the kinetic energy loss due to the phase winding of $A_{++}$
with $w_{++}=1$, which remains unscreened and spreads out whole system,
becomes less compared to AT at rest, meaning that the HQV
energy approaches towards the AT energy in Fig. 1 at $\Omega=0$ as seen
from inset of Fig. 3.
Under rotation the HQV energy decreases by absorbing the
rotation kinetic energy and eventually becomes lower at $\Omega_{c1}$,
which is smaller than $\Omega_{c2}$, stabilizing HQV over SV.

The main panel in Fig. 3 shows $\Omega_{c1}$ and $\Omega_{c2}$ as a
function of the system size $R$. It is seen that
the relative stability region $\Omega_{c1}/\Omega_{c2}\sim0.8$
stays at a constant against $R$, keeping 20$\%$ region
above the critical rotation speed $\Omega_{c1}$
at which single HQV is created in the system.
The extrapolated $\Omega_{c1}$ to $R$=1.5mm,
by which Yamashita, et al\cite{yamashita} have performed experiments,
amounts to $\Omega_{c1}\sim 0.05 $rad/sec.
The rotation speed stability of the rotation cryostat at ISSP,
Univ. Tokyo is accurate enough to perform it.
We also notice that by changing the radius $R$ of the system one
can control the $\Omega_{c1}$ value at will.
For example, in $R=300\mu$m,  $\Omega_{c1} \sim 1$rad/sec which
is convenient speed to run their rotating cryostat.

\begin{figure}
\includegraphics[width=8cm]{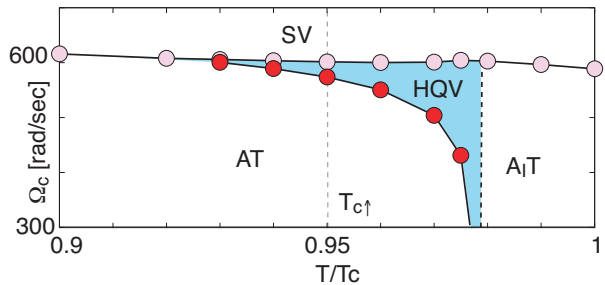}
\caption{(Color online)
Stability region of HQV in $\Omega$ versus $T/T_c$
($R=10\mu$m and $\Delta t=0.05$). A$_1$T
denotes the A$_1$ phase texture where only
$\downarrow\downarrow$ pairs exist.
}
\end{figure}

In Fig. 4 we depict the temperature dependence of the HQV stability region.
We see that the stability region for HQV is widen divergently when
approaching the lower critical temperature $T_{c\uparrow}$ from below
where the disparity of the OP amplitudes between $\uparrow\uparrow$ pair
and  $\downarrow\downarrow$ pair increases.
Note that the actual lower transition temperature $T_{c\uparrow}$
is shifted slightly upward because the spatially varying
$\uparrow\uparrow$ pair OP $A_{-+}(r)$ induces  $A_{++}(r)$.
In other words, the $A_1$ phase for $T_{c\downarrow}>T>T_{c\uparrow}$
becomes narrower. Thus one needs not only careful temperature
control, an order of 0.01K which is feasible enough, but also
theoretical backup to estimate this shift in order to precisely locate
the HQV stability region under actual experimental setup.



Since the HQV has the odd winding number for the $\uparrow\uparrow$
pairs, the Majorana quasi-particle with zero energy exactly
at the Fermi level, which is localized in the vortex core,
is ensured by both the index theorem based on topological
argument\cite{tewari}, or directly solving the
Bogoliubov-de Gennes equation \cite{tsutsumi}.
These arguments are based on the assumption that
the $\uparrow\uparrow$ pair and $\downarrow\downarrow$ pair
are completely decoupled.
Here the situation is more subtle.
The $\uparrow\uparrow$ pair and $\downarrow\downarrow$ pair are interacting through
the fourth order GL terms, which is mentioned above.
These terms comes from the strong coupling effect due to ferromagnetic
spin fluctuations\cite{anderson}, which ultimately
help stabilizing the present HQV. Therefore, it is not obvious completely
that the present HQV can accommodate the Mojorana fermion with the
exactly zero energy. This issue belongs to a future problem.

We also remark on the experimental point that the identification
of the HQV is not an easy task.
The HQV and SV are indistinguishable by the usual
NMR method which utilizes the satellite position in
the spectrum\cite{yamashita} because $\vec d\perp \vec l$
are always kept for both vortices, giving rise to the identical
NMR spectra. We suggest small tilting of the field direction
from $H\parallel z$ may yield the different NMR signatures.
This point deserves further elaboration.

Finally it should be pointed out that our previous theory for the
parallel geometry of the superfluid $^3$He\cite{tsutsumi} differs in the field orientation $H\perp z$ there. The singular vortex with odd
integer winding number was found in this spinless chiral superfluid. This
also gives rise to the Majorana zero energy mode.
Thus the field orientations yield different vortices, but
those accommodate the Majorana particle localized at each vortex core.

In conclusion, we have found the stability region of half quantum
vortex in $T$-$\Omega$ plane of superfluid $^3$He A phase
confined in parallel plates and given physical reasons why it is
stabler than A phase texture or ordinary singular vortex.
We propose a concrete experimental setup, which is feasible
by using the rotating cryostat such as in ISSP, Univ. Tokyo.

We thank T. Ohmi, O. Ishikawa, M. Yamashita, R. Ishiguro,
K. Izumina, M. Kubota, T. Mizushima, M. Ichioka, and G.E. Volovik for useful discussions.


\begin{thebibliography}{99}
\bibitem{GLparameters} The following GL parameters are
used: $\alpha_0=3.81\times10^{50}[J^{-1}m^{-3}]$, and $\beta_1=-3.75, 
\beta_2=6.65,\beta_3=6.56,\beta_4=5.99, \beta_5=-8.53$ in units of $10^{99}[J^{-3}m^{-3}].$
\end{thebibliography}
\end{document}